\documentstyle[11pt,epsfig]{article}
\begin{document}
\topskip 2cm
\begin{center} {\large\bf Master Equations for Master Amplitudes $^{\star}$ } 
\end{center} 
\vspace{1.2cm} 
\begin{center} 
{ {\large M.~Caffo$^{ab}$, H.~Czy{\.z}\ $^{c}$, S.~Laporta$^{b}$} 
and {\large E.~Remiddi$^{ba}$ \\ } }
\end{center} 
\begin{itemize} 
\item[$^a$] {\sl INFN, Sezione di Bologna, I-40126 Bologna, Italy } 
\item[$^b$] {\sl Dipartimento di Fisica, Universit\`a di Bologna, 
                 I-40126 Bologna, Italy } 
\item[$^c$] {\sl Institute of Physics, University of Silesia, 
                 PL-40007 Katowice, Poland } 
\end{itemize} 
e-mail: {\tt caffo@bo.infn.it \\
\hspace*{1.1cm} czyz@usctoux1.cto.us.edu.pl \\
\hspace*{1.1cm} laporta@bo.infn.it \\
\hspace*{1.1cm} remiddi@bo.infn.it \\ }
\begin{center} \begin{abstract} 
The general lines of the derivation and the main properties of the master 
equations for the master amplitudes associated to a given Feynman graph 
are recalled. Some results for the 2-loop self-mass graph with 4 propagators 
are then presented. 
\end{abstract} \end{center} 
{\scriptsize \noindent ------------------------------- \\
PACS 11.10.-z Field theory \\
PACS 11.10.Kk Field theories in dimensions other than four \\
PACS 11.15.Bt General properties of perturbation theory    \\ } 
\vfill 
\footnoterule \noindent 
$^{\star}${\footnotesize Presented at the Zeuthen Workshop on Elementary 
           Particle Physics - Loops and Legs in Gauge Theories - 
           Rheinsberg, 19-24 April 1998. } 
\pagestyle{plain} \pagenumbering{arabic} 
\def\Li2{\hbox{Li}_2} 
\def\LLL{L(m_1^2,m_2^2,m_3^2)} 
\def\a{\alpha} 
\def\app{{\left(\frac{\alpha}{\pi}\right)}} 
\newcommand{\Eq}[1]{Eq.(\ref{#1})} 
\newcommand{\labbel}[1]{\label{#1}} 
\newcommand{\cita}[1]{\cite{#1}} 
\newcommand{\dnk}[1]{ \frac{d^nk_{#1}}{(2\pi)^{n-2}} } 
\newcommand{\e}{{\mathrm{e}}} 
\newcommand{\om}{\omega} 
\newcommand{\verso}[1]{ {\; \buildrel {n \to #1} \over{\longrightarrow}}\; } 
\newcommand{\F}[1]{F_{#1}(n,m_1^2,m_2^2,m_3^2,p^2)} 
\newcommand{\G}{G(n,m_1^2,m_2^2,m_3^2,m_4^2,p^2)} 
\newcommand{\Sm}{S(n,m_1^2,m_2^2,p^2)} 
\newcommand{\Rp}{R^2(-p^2,m_1^2,m_4^2)} 
\section{ Introduction. } 
The integration by part identities \cita{CT} are by now a standard tool 
for obtaining relations between the many integrals associated to any 
Feynman graph or, equivalently, for working out recurrence relations for 
expressing the generic integral in terms of the "master integrals" 
or ``master amplitudes" of the considered graph. A good example of the 
use of the integration by part identities is given in \cita{Tarasov}, 
where the recurrence relations for all the 2-loop self-mass amplitudes 
are established in the arbitrary masses case. \\ 
It has been shown in 
\cita{ER} that by that same technique one can obtain a set of linear 
first order differential equations for the master integrals themselves; 
the coefficients of the equations are ratio of polynomials with integer 
coefficients in all the variables; the equations are further non homogeneous, 
with the non homogeneous terms given by the master integrals of the 
simpler graphs obtained from the considered graph by removing one or 
more internal propagators. \\ 
Restricting ourselves for simplicity to the self-mass case, for any 
Feynman graph the related integrals can in general be written in the form 
\begin{equation} 
   A(\alpha,p^2) = \int d^nk \ B(\alpha,p,k) \ . 
\labbel{1} \end{equation} 
In more detail, \( d^nk = d^nk_1...d^nk_l \) stands for the 
\( n \)-continuous integration 
on an arbitrary number \( l \) of loops and \( k \) stands for the set 
of the corresponding loop momenta, so that there are altogether 
\( s=(l+1)(l+2)/2 \) different scalar products, including \( p^2 \); 
\( B(\alpha,p,k) \) is the product of any power of the scalar products 
in the numerators divided by any power of the propagators occurring in the 
graph (all masses will always be taken as different, unless otherwise 
stated); as the propagators are also simple combinations of the scalar 
products, simplifications might occur between numerator and denominator 
and as a consequence one expects quite in general \( (s-1) \) different 
factors altogether in the numerator and denominator, independently of 
the actual number of propagators present in the graph (graphs with less 
propagators have more factors in the numerator and {\it viceversa}); 
therefore the symbol \( \alpha \) in \Eq{1} stands in fact for a set of 
\( (s-1) \) indices -- the (integer) powers of the \( (s-1) \) factors. \\ 
The integration by parts corresponding to the amplitudes of \Eq{1} are 
\begin{equation} 
   \int d^nk \frac{\partial}{\partial k_{i,\mu}} 
        \Bigl[ v_\mu B(\alpha,p,k) \Bigr] = 0 \ , \hskip 1cm i=1,...,l 
\labbel{2} \end{equation} 
where \( v \) stands for any of the \( (l+1) \) vectors \( k \) and \( p \); 
there are therefore \( l(l+1) \) identities for each set of indices 
\( \alpha \). The identity is easily established -- for small \( n \) the 
integral of the divergence vanishes. When the derivatives are explicitly 
carried out, one obtains the sum of a number of terms, all equal to a 
simple coefficient (an integer number or, occasionally, \( n \)), times 
an integrand of the form \( B(\beta,k,p) \), with the set of indices 
\( \beta \) differing at most by a unity in two places from the set 
\( \alpha \). \\ 
That set of identities is infinite; even if they are not all 
independent, they can be used for obtaining the recurrence relations, 
by which one can express each integral in terms of a few already 
mentioned ``master amplitudes", through a relation of the form 
\begin{equation} 
   A(\alpha,p^2) = \sum \limits_{m} C(\alpha,m) A(m,p^2) 
                 + \sum \limits_{j} C(\alpha,j) A(j,p^2) \ , 
\labbel{3} \end{equation} 
where the set of indices \( m \) takes the very few values corresponding 
to the master amplitudes, \( j \) refers to simpler master integrals 
in which one or more denominators are missing, and the coefficients 
\( C(\alpha,m), C(\alpha,j) \) are ratios of polynomials in \( n \), 
masses and \( p^2 \). \\ 
Let us consider now one of the master amplitudes themselves, say the 
master amplitude identified by the set of indices \( m \); according 
to \Eq{1} we can write 
\begin{equation} 
   A(m,p^2) = \int d^nk \ B(m,p,k) \ . 
\labbel{4} \end{equation} 
By acting with \( p_\mu (\partial/\partial p_\mu) \) on both sides we get 
\begin{equation} 
   p^2 \frac{\partial}{\partial p^2} A(m,p^2) = \frac{1}{2} 
       \int d^nk \ p_\mu \frac{\partial}{\partial p_\mu} \ B(m,p,k) \ . 
\labbel{5} \end{equation} 
According to the discussion following \Eq{2}, the {\it r.h.s.} is a 
combination of integrands; as \Eq{3} applies to each of the corresponding 
integrals, one obtains the relations 
\begin{equation} 
   p^2 \frac{\partial}{\partial p^2} A(m,p^2) = 
                   \sum \limits_{m'} C(m,m') A(m',p^2) 
                 + \sum \limits_{j}  C(m,j)  A(j,p^2) \ , 
\labbel{6} \end{equation} 
which are the required master equations. As in \Eq{3}, \( j \) refers 
to simpler master integrals (in which one or more denominators are 
missing; they constitute the non-homogeneous part of the master 
equations), to be considered as known when studying the \( A(m,p^2) \). \\ 
It is obvious from the derivation that the master equations can be 
established regardless of the number of loops. It is equally clear that 
for graphs depending on several external momenta  (such as vertex or 
4-body scattering graphs) one has simply to replace the single operator 
\( p_\mu (\partial/\partial p_\mu) \) of \Eq{5} by the set of operators 
\( p^i_\mu (\partial/\partial p^j_\mu) \), where \( i,j \) run on all the 
external momenta, and with some more algebra one can obtain master 
equations in any desired Mandelstam variable. \\ 
The master equations are a powerful tool for the study and the evaluation 
of the master amplitudes; among other things: 
\begin{itemize} 
\item they provide information on the values of the master amplitudes at 
      special kinematical points (such as \( p^2=0 \) in \Eq{6}; the 
      {\it l.h.s.} vanishes, as \( p^2=0 \) is a regular point, so that 
      the {\it r.h.s.} is a relation among master amplitudes at \( p^2=0 \), 
      usually sufficient to fix their values at that point); 
\item the master equations are valid identically in \( n \), so that they 
      can be expanded in \( (n-4) \) and solved recursively for the various 
      terms of the expansion in \( (n-4) \), starting from the most 
      singular (with  2-loop amplitudes one expects at most a double 
      pole in \( (n-4) \)); 
\item when the initial value at \( p^2 = 0 \) has been obtained, the 
      equations can be integrated by means of fast and precise numerical 
      methods (for instance with a Runge-Kutta routine), so providing a 
      convenient approach to their numerical evaluation; note that the 
      numerical approach can be used both for arbitrary \( n \) or for 
      \( n=4 \), once the expansion has been properly carried out; 
\item the equations can be used to work out virtually any kind of 
      expansion, in particular the large \( p^2 \) expansion, as will 
      be shown in some detail; 
\item in particularly simple cases (for instance when most of the masses 
      vanish and only one or two scales are left) the analytic quadrature 
      of the equations can lead to the analytic evaluation of the master 
      amplitudes.       
\end{itemize} 
\section{The 2-loop 4-propagator graph.} 
The use of the master equations for studying the 1-loop self-mass and 
the 2-loop sunrise self-mass graph has been already discussed, \cita{ER}, 
\cita{CCLR}; we will describe here its application to the 
2-loop 4-propagator self-mass graph shown in Fig.1. \\ 

\begin{figure}[h]
\epsfbox[0 40 140 140]{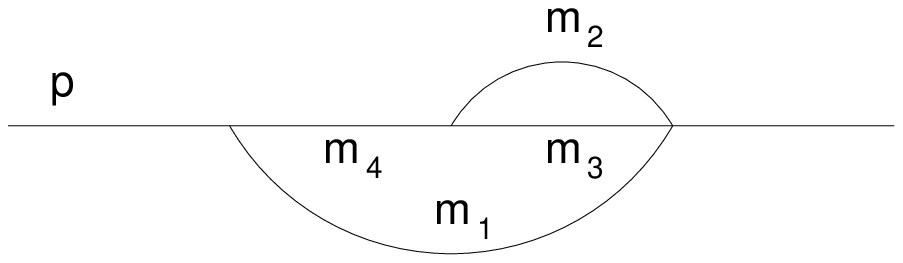} 
\vspace{-8mm}
\caption{\( \G \), the 2-loop 4-propagator self-mass graph.} 
\end{figure}
\vspace{\baselineskip} 

\noindent The corresponding amplitude is defined as 
\begin{eqnarray} 
  \G &=& \frac{1}{C^2(n)} \int \frac{d^nk_1}{(2\pi)^{n-2}} \ 
                          \frac{d^nk_2}{(2\pi)^{n-2}} \nonumber \\ 
    && {\kern-105pt} \frac{1} 
  {(k_1^2+m_1^2)\ [(p-k_1)^2+m_4^2]\ (k_2^2+m_2^2)\ [(p-k_1-k_2)^2+m_3^2]} ; 
\labbel{6a} \end{eqnarray} 
the conventional factor \( C(n) \) is defined in \cita{CCLR}, it is 
sufficient to know that at \( n=4 \) its value is 1; if all momenta are 
Euclidean the \( i\epsilon \) in the propagators is not needed. \\
Skipping details, the master equation is 
\\ \vbox{ \begin{eqnarray} 
  && {\kern-30pt} \Rp p^2 \frac{\partial}{\partial p^2} \G = \nonumber \\ 
  && \phantom{+} \frac{n-4}{2} \Rp \G                      \nonumber \\ 
  && + (n-3)\left[ (m_1^2+m_4^2)p^2 + (m_1^2-m_4^2)^2 \right] \G 
                                                           \nonumber \\ 
  && + (3p^2-m_1^2+m_4^2) m_1^2 \F{1}         \nonumber \\ 
  && + (p^2-m_1^2+m_4^2) \Biggl[ \frac{3n-8}{2}  \F{0}     \nonumber \\ 
  && {\kern+30pt} + m_2^2 \F{2} + m_3^2 \F{3}           \nonumber \\ 
  && {\kern+30pt} - \frac{1}{2}(n-2) V(n,m_2^2,m_3^2,m_4^2) \Biggr] 
\labbel{7} \end{eqnarray} } 
where the \( \F{k}, k=0,..,3 \) are the 2-loop self-mass sunrise master 
amplitudes \cita{Tarasov}\cita{CCLR}, 
\begin{eqnarray} 
    \F{0}     
    &=& \frac{1}{C^2(n)} \int \frac{d^nk_1}{(2\pi)^{n-2}} \ 
                          \frac{d^nk_2}{(2\pi)^{n-2}} \nonumber \\ 
    && {\kern-75pt} \frac{1} 
      {(k_1^2+m_1^2)\ (k_2^2+m_2^2)\ [(p-k_1-k_2)^2+m_3^2]} ; 
\labbel{6b} \end{eqnarray} 
and 
\begin{equation} 
   \F{i} = - \frac{\partial}{\partial m_i^2} \F{0},\ \ i=1,2,3 
\labbel{6c} \end{equation} 
while \( V(n,m_1^2,m_2^2,m_3^2) \) corresponds to the 2-loop vacuum 
amplitude, 
\begin{eqnarray} 
    V(n,m_1^2,m_2^2,m_3^2) 
    &=& \frac{1}{C^2(n)} \int \frac{d^nk_1}{(2\pi)^{n-2}} \ 
                          \frac{d^nk_2}{(2\pi)^{n-2}} \nonumber \\ 
    && {\kern-50pt} \frac{1} 
      {(k_1^2+m_1^2)\ (k_2^2+m_2^2)\ [(k_1+k_2)^2+m_3^2]} 
\labbel{6d} \end{eqnarray} 
and, as usual,  
\[ R^2(-p^2,m_1^2,m_2^2) = 
                    p^4+m_1^4+m_2^4+2m_1^2p^2+2m_2^2p^2-2m_1^2m_2^2 \ . \] 
\noindent The value at \( p^2 = 0 \) is (almost trivially) found to be 
\begin{equation} 
 G(n,m_1^2,m_2^2,m_3^2,m_4^2,0) = 
    \frac{V(n,m_2^2,m_3^2,m_4^2)-V(n,m_1^2,m_2^2,m_3^2)}{m_1^2-m_4^2} 
\labbel{8} \end{equation} 
The expansion in \( (n-4) \) reads 
\begin{equation} 
  \G = \sum \limits_{k=-2}^{\infty} 
       (n-4)^k G^{(k)}(m_1^2,m_2^2,m_3^2,m_4^2,p^2) 
\labbel{9a} \end{equation} 
By expanding in the same way all the other amplitudes occurring in the 
master equation \Eq{7} and using the results of \cita{CCLR}, 
the first values are found to be 
\begin{eqnarray} 
    G^{(-2)}(m_1^2,m_2^2,m_3^2,m_4^2,p^2) &=& + \frac{1}{8} \nonumber \\ 
    G^{(-1)}(m_1^2,m_2^2,m_3^2,m_4^2,p^2) &=& - \frac{1}{16}  
                      - \frac{1}{2} S^{(0)}(m_1^2,m_4^2,p^2) \ , 
\labbel{9} \end{eqnarray} 
where \( S^{(0)}(m_1^2,m_4^2,p^2) \) is the finite part at \( n=4 \) of 
the 1-loop self mass; more exactly, defining 
\[ \Sm = \frac{1}{C^(n)} \int \frac{d^nk}{(2\pi)^{n-2}} \ 
    \frac{1} {(k^2+m_1^2)\ [(p-k)^2+m_2^2]} \ ,                       \] 
and expanding in \( (n-4) \), one finds \cita{ER} 
\begin{equation} 
  \Sm = - \frac{1}{2} \frac{1}{(n-4)} + S^{(0)}(m_1^2,m_2^2,p^2) 
    + {\cal O} (n-4) 
\labbel{12} \end{equation} 
with 
\begin{eqnarray} 
  S^{(0)}(m_1^2,m_2^2,p^2) &=& \frac{1}{2} - \frac{1}{4} \ln(m_1m_2) 
                                             \nonumber \\ 
   && {\kern-80pt} + \frac{1}{4p^2} \Biggl[ R(-p^2,m_1^2,m_2^2) 
   \ln(u(p^2,m_1^2,m_2^2)) + (m_1^2-m_2^2)\ln{\frac{m_1}{m_2}} \Biggl] 
\labbel{13} \end{eqnarray} 
where 
\[ R(-p^2,m_1^2,m_2^2) = \sqrt{ [p^2+(m_1+m_2)^2] [p^2+(m_1-m_2)^2] } \] 
\[ u(p^2,m_1^2,m_2^2) = \frac 
     { \sqrt{ p^2+(m_1+m_2)^2 } - \sqrt{ p^2+(m_1-m_2)^2 } } 
     { \sqrt{ p^2+(m_1+m_2)^2 } + \sqrt{ p^2+(m_1-m_2)^2 } } \ . \] 
\section{The large \( p^2 \) expansion.} 
Quite in general, if {\kern+5pt} \( \om = (n-4)/2 \) , the large 
\( p^2 \) expansion is 
\begin{eqnarray} 
  \G &=& (p^2)^{2\om} \sum \limits_{k=0}^{\infty} 
          G_k^{(\infty,2)}(n,m_1^2,m_2^2,m_3^2,m_4^2) 
                                      \frac{1}{(p^2)^k} \nonumber \\ 
    &+& (p^2)^{\om} \sum \limits_{k=0}^{\infty} 
          G_k^{(\infty,1)}(n,m_1^2,m_2^2,m_3^2,m_4^2) 
                                      \frac{1}{(p^2)^k} \nonumber \\ 
    &+& \frac{1}{p^2} \sum \limits_{k=0}^{\infty} 
          G_k^{(\infty,0)}(n,m_1^2,m_2^2,m_3^2,m_4^2) 
                                      \frac{1}{(p^2)^k} 
\labbel{14} \end{eqnarray} 
Any \( l \)-loop amplitude, indeed, at large \( p^2 \) develops \( l \) terms 
with ``fractional powers" in \( p^2 \), with exponents 
\( \om, 2\om, \cdots l\om \), besides the ``regular" term 
containing integer powers only. 
As any 2-loop amplitude has ``fractional dimension" equal to 2 (in square 
mass units), on dimensional grounds the coefficients \( G_k^{(\infty,2)}, 
G_k^{(\infty,1)} \) and \( G_k^{(\infty,0)} \) must have ``fractional 
dimension" equal to 0, 1 and 2 respectively. \\ 
Similar expansions are valid for the sunrise amplitudes appearing in the 
{\it r.h.s.} of \Eq{7} such as 
\begin{eqnarray} 
  F_0 (n,m_1^2,m_2^2,m_3^2,p^2) &=& 
    p^2 (p^2)^{2\om} \sum \limits_{k=0}^{\infty} 
          F_{0,k}^{(\infty,2)}(n,m_1^2,m_2^2,m_3^2) 
                                      \frac{1}{(p^2)^k} \nonumber \\ 
    &+& (p^2)^{\om} \sum \limits_{k=0}^{\infty} 
          F_{0,k}^{(\infty,1)}(n,m_1^2,m_2^2,m_3^2) 
                                      \frac{1}{(p^2)^k} \nonumber \\ 
    &+& \frac{1}{p^2} \sum \limits_{k=0}^{\infty} 
          F_{0,k}^{(\infty,0)}(n,m_1^2,m_2^2,m_3^2) 
                                      \frac{1}{(p^2)^k} 
\labbel{15} \end{eqnarray} 
as well as for the 1-loop self-mass amplitude (whose ``fractional power" 
is just \( \om \)) 
\begin{eqnarray} 
  S (n,m_1^2,m_2^2,p^2) &=& 
        (p^2)^{\om} \sum \limits_{k=0}^{\infty} 
        S_k^{(\infty,1)}(n,m_1^2,m_2^2) \frac{1}{(p^2)^k} \nonumber \\ 
    &+& \frac{1}{p^2} \sum \limits_{k=0}^{\infty} 
        S_k^{(\infty,0)}(n,m_1^2,m_2^2) \frac{1}{(p^2)^k} \ . 
\labbel{16} \end{eqnarray} 
When the above expansions are inserted into the master equations and in 
the recurrence relations (to be regarded, in this context, as differential 
equations in the masses), one obtains a number of equations providing 
important relations between the coefficients of the expansions. \\ 
As a result, one finds \cita{ER} 
\begin{equation} 
  S_0^{(\infty,1)}(n,m_1^2,m_2^2) = S^{(\infty)}(n) \ , 
\labbel{17} \end{equation} 
where \( S^{(\infty)}(n) \) is a dimensionless function of \( n \) only; 
an explicit calculation (most easily performed by putting \( m_2 = 0 \))  
gives for its expansion in \( n-4 \) 
\begin{equation} 
 S^{(\infty)}(n) = - \frac{1}{2} \ \frac{1}{n-4} + \frac{1}{2} 
                   + \left( \frac{1}{8} \zeta(2) - \frac{1}{2} \right) (n-4) 
                   + {\cal O} \left( (n-4)^2 \right) . 
\labbel{18} \end{equation} 
All the other \( S_k^{(\infty,1)}(n,m_1^2,m_2^2), k=1,2,.. \) can then 
obtained explicitly and are found to be proportional to 
\( S^{(\infty)}(n) \). \\ 
One further finds 
\begin{equation} 
  S_0^{(\infty,0)}(n,m_1^2,m_2^2) = \frac { m_1^{n-2} + m_2^{n-2} } 
                                          {(n-2)(n-4)} 
\labbel{19} \end{equation} 
and similar expressions for all the other \( S_k^{(\infty,0)} \). \\ 
Note here that in the massless limit one has the exact relation 
\begin{equation} 
   S(n,0,0,p^2) = (p^2)^\om S^{(\infty)}(n) \ . 
\labbel{20} \end{equation} 
Likewise, one obtains (\cita{CCLR}; the results are reported here with 
minor changes of notation) 
\begin{equation} 
    F_{0,0}^{(\infty,2)}(n,m_1^2,m_2^2,m_3^2) = F^{(\infty,2)}(n) 
\labbel{21} \end{equation} 
\begin{equation} 
    F_{0,0}^{(\infty,1)}(n,m_1^2,m_2^2,m_3^2) = F^{(\infty,1)}(n) 
                        \left( m_1^{n-2} + m_2^{n-2} + m_3^{n-2} \right) 
\labbel{22} \end{equation} 
\begin{equation} 
    F_{0,0}^{(\infty,0)}(n,m_1^2,m_2^2,m_3^2) = 
       \frac{ (m_1 m_2)^{n-2} + (m_1 m_3)^{n-2} + (m_2 m_3)^{n-2} } 
            {(n-2)^2(n-4)^2} 
\labbel{23} \end{equation} 
By expanding as usual around \( n=4 \) 
\begin{equation} 
    F^{(\infty,i)}(n) = \sum \limits_{j=-2}^{\infty} (n-4)^j F^{(i,j)} \ , 
\labbel{24} \end{equation} 
one finds 
\begin{eqnarray} 
  F^{(2,-2)} &&{\kern-20pt}= 0 \ \ , \ \ \ \ \ \ \ \ \ \ \ \
  F^{(1,-2)} = -\frac{1}{4}   \nonumber \\
  F^{(2,-1)} &&{\kern-20pt}= \frac{1}{32}    \ \ , \ \ \ \ \
      \ \ \ \ \
  F^{(1,-1)} = \frac{3}{8}     \nonumber \\
  F^{(2,0)} &&{\kern-20pt}= -\frac{13}{128} \ \ , \ \ \ \ \ \
  F^{(1,0)\phantom{-}} = - 4F^{(2,1)} + \frac{59}{128} \ .
\labbel{25} \end{eqnarray} 
When the large \( p^2 \) expansions, Eq.s(\ref{14},\ref{15},\ref{16}), 
are substituted into \Eq{7}, one can express the 
\( G_k^{(\infty,i)}(n,m_1^2,m_2^2,m_3^2,m_4^2) \) in terms of the other 
coefficients, already known. \\ 
One finds 
\begin{equation} 
   G_0^{(\infty,0)}(n,m_1^2,m_2^2,m_3^2,m_4^2) = V(n,m_2^2,m_3^2,m_4^2) 
\labbel{26} \end{equation} 
and 
\begin{equation} 
   G_0^{(\infty,2)}(n,m_1^2,m_2^2,m_3^2,m_4^2) = 
     \frac{3n-8}{n-4} F^{(\infty,2)}(n) \ . 
\labbel{27} \end{equation} 
while \( G_0^{(\infty,1)}(n,m_1^2,m_2^2,m_3^2,m_4^2) \) depends on the 
combination 
\[ (n-2) F^{(\infty,1)}(n) - S^{(\infty)}(n)/(n-4) \ ; \] 
when the combination is expanded around \( n=4 \) as in \Eq{24} and 
the explicit values of \Eq{25} are used, the first 3 terms of the 
expansion -- namely the double pole, the simple pole and the term constant 
in \( (n-4) \) -- are all found to vanish, suggesting the existence of the 
exact relation 
\begin{equation} 
   F^{(\infty,1)}(n) = \frac{1}{(n-2)(n-4)} S^{(\infty)}(n) \ . 
\labbel{28} \end{equation} 
The above result seems to be confirmed by a preliminary investigation of 
the large \( p^2 \) behaviour of the 5-propagator 2-loop self-mass graph 
(in progress). When \Eq{28} is taken as valid, one finds 
\begin{equation} 
   G_0^{(\infty,1)}(n,m_1^2,m_2^2,m_3^2,m_4^2) = 0 
\labbel{29} \end{equation} 
and 
\begin{eqnarray} 
   G_1^{(\infty,1)}(n,m_1^2,m_2^2,m_3^2,m_4^2) &=& \nonumber \\ 
   && {\kern-65pt}           \frac{S^{(\infty)}(n)}{(n-2)(n-4)} 
          \Bigl[ (m_1^2)^\om - (n-3) \left( (m_2^2)^\om 
                                    + (m_3^2)^\om \right) \Bigr] 
\labbel{30} \end{eqnarray} 
\vskip 2 truecm \noindent 
{\large\bf Acknowledgements.} 
As in previous work, the algebra needed in all the steps of the work 
has been processed by means of the computer program {\tt FORM} 
\cita{FORM} by J. Vermaseren. One of the authors (E.R.) is glad to 
acknowledge an interesting discussion with K.G. Chetyrkin on the 
universality of the coefficients of the large \( p^2 \) expansions. 
\vskip 2 truecm 
\def\NC{{\sl Nuovo Cimento }\ } 
\def\NP{{\sl Nuc. Phys. }\ } 
\def\PL{{\sl Phys. Lett .}\ } 
\def\PR{{\sl Phys. Rev. }\ } 
\def\PRL{{\sl Phys. Rev. Lett. }\ } 

\end{document}